\documentclass[%
reprint,
amsmath,amssymb,
aps,
]{revtex4-2}
\usepackage[colorlinks=true,linkcolor=blue,citecolor=blue,urlcolor=
blue]{hyperref}

\usepackage{graphicx}
\usepackage{dcolumn}
\usepackage{bm}
\usepackage{amsmath}
\usepackage[dvipsnames]{xcolor}

\begin{document}
	
	\preprint{APS/123-QED}
	
	\title{Relativistic resistive dissipative magnetohydrodynamics from the relaxation time approximation}
	
	\author{Ankit Kumar Panda}
	\email{ankitkumar.panda@niser.ac.in}
	\author{Ashutosh Dash}%
	\email{ashutosh.dash@niser.ac.in}
	\author{Rajesh Biswas}%
	\email{rajeshbiswas@niser.ac.in}
	\author{Victor Roy}%
	\email{victor@niser.ac.in}
	\affiliation{National Institute of Science Education and Research, Bhubaneswar, HBNI, Jatni, 752050, India}%

	\date{\today}
	
	\begin{abstract}
		Here we derive the relativistic resistive dissipative second-order magnetohydrodynamic evolution equations using the Boltzmann equation,  thus extending our work from the previous paper \href{https://link.springer.com/article/10.1007/JHEP03(2021)216}{JHEP 03 (2021) 216} where we considered the non-resistive limit. We solve the Boltzmann equation for a system of particles and antiparticles using the relaxation time approximation and the Chapman-Enskog like gradient expansion for the off-equilibrium distribution function, truncating beyond second-order. 
In the first order, the bulk and shear stress are independent of the electromagnetic field, however, the diffusion current, shows a dependence on the electric field.
In the second-order, the new transport coefficients that couple electromagnetic field with the dissipative quantities appear, which are different from those obtained in the 14-moment approximation~\cite{Denicol:2019iyh} in the presence of the electromagnetic field. Also we found out the various components of conductivity in this case.
	\end{abstract}
	
	\maketitle
	
	
	\section{\label{sec:intro}Introduction}

	The dynamical evolution of hot and dense nuclear matter produced in high-energy heavy-ion collisions has been
	successfully described by the causal second-order viscous hydrodynamics numerical solution~\cite{Rischke:1998fq,Shen:2011zc,Luzum:2008cw,Heinz:2013th,Bozek:2012qs,Roy:2012jb,Heinz:2011kt,Niemi:2012ry,Schenke:2011bn,Hirano:2005xf}. 
	The discovery that the Quark-Gluon-Plasma (QGP) produced in such high-energy heavy-ion collisions is a near-perfect fluid is primarily
	based on phenomenological studies using relativistic viscous hydrodynamics~\cite{Chaudhuri:2012yt,Gale:2013da,Jeon:2015dfa,Romatschke:2017ejr} . 
	However, almost all of these studies
	neglected the possible effects of strong transient electromagnetic fields produced in the initial stage of high-energy
	heavy-ion collisions. The finite electrical conductivity of the QGP and the ambient intense electromagnetic fields
	strongly suggest that the most appropriate framework for this case is relativistic resistive viscous magnetohydrodynamics.
	In our previous work~\cite{Panda:2020zhr}, we derived the second-order causal relativistic ideal viscous magnetohydrodynamics (MHD)
	equations from the Relaxation Time Approximation (RTA). Here we extend our previous work to include the finite
	resistivity of the fluid. Note that in Ref.~\cite{Denicol:2019iyh} the formulation for resistive MHD was derived for the first time from the
	moment method. It was shown for the ideal-MHD case that although the RTA~\cite{Panda:2020zhr} and moment methods~\cite{Denicol:2018rbw} give similar evolution equations for dissipative stresses, the two formulations give different values of transport
	coefficients. Moreover, we also showed that the RTA formulation gives rise to some new transport coefficients.

	It is worthwhile to mention that ideal-MHD (infinite electrical conductivity $\sigma$) limit is an approximation works
	only in limited systems. As was pointed out in Ref.~\cite{Denicol:2019iyh} that this approximation has a basic flaw in the sense that
	$\sigma$ is a transport coefficient, and like other transport coefficients (e.g., shear and bulk viscosity) is proportional to
	the mean free path of the microscopic degrees of freedom. It is inconsistent to take $\sigma \rightarrow \infty$
	while other transport coefficients remain finite (to be precise the magnetic Reynolds number governs the ideal/resistive regime). In a resistive fluid the magnetic field can generally move through the fluid
	following a diffusion law with the resistivity of the plasma serving as a diffusion constant. This implies that the ideal
	MHD approximation is only good for a given length and time scale before the diffusion becomes non-negligible.

	The resistive MHD allows finite electric field inside the plasma while in the ideal MHD the electric field is constrained
	via $\bf{E}=-\bf{v}\times \bf{B}$. Since charged particle motion in cross electric and magnetic field becomes much
	more complicated than only magnetic field case, the resistive-MHD consequently shows much more complex
	behaviour (e.g., magnetic reconnection) than ideal-MHD. In addition to the regular applications in solar and cosmological
	systems the RMHD  has recently found applications in condensed matter systems such as  Dirac~\cite{crossno} and Weyl semimetal~\cite{Lucas:2015sya}. In heavy ion collisions, the importance of RMHD has recently been realised in Refs.~\cite{Roy:2015kma,Pu:2016ayh,Siddique:2019gqh,Pu:2016bxy,Shokri:2018qcu,Moghaddam:2017myy,Inghirami:2019mkc} and it is an active area of research specifically for detecting CME~\cite{Kharzeev:2007jp,Fukushima:2008xe,Kharzeev:2015znc} and other related phenomena.
	
Throughout the paper we use the natural units, $\hbar=c=k_{B}=\mu_{0} =\epsilon_{0}=1$ and the metric tensor used is $g^{\mu\nu}=$diag$\left(+1,-1,-1,-1\right)$. That means the current formalism is applicable only for the flat-space time. We also use the following definitions: $u \cdot p= u^{\mu}p_{\mu}$ , $\Delta^{\mu\nu}=g^{\mu\nu}-u^{\mu}u^{\nu}$ , where $u^{\mu}=\gamma(1,\mathbf{v})$ is the fluid four-velocity and $\gamma=(1-\mathbf{v}^2)^{-1/2}$. In the rest frame of the fluid $u^{\mu}=\left(1,\mathbf{0}\right)$. The co-moving derivative is given by $u^{\mu}\partial_{\mu}=D$ (sometimes we use overdot to denote the comoving derivative), and the partial derivative can be decomposed into a comoving and a spatial part (in the rest frame) as $\partial_{\mu}=u_{\mu}D+\nabla_{\mu}$, where $\nabla_{\mu} \equiv \Delta^{\gamma}_{\mu}\partial_{\gamma}$. We also use the following decomposition
	\begin{equation}
		\label{eq:nablaanduexpansion}
		\nabla^{\alpha}u^{\beta}=\omega^{\alpha\beta}+\sigma^{\alpha\beta}+\frac{1}{3}\theta\Delta^{\alpha\beta},
	\end{equation}
	where $\omega^{\alpha\beta}=(\nabla^{\alpha}u^{\beta}-\nabla^{\beta}u^{\alpha})/2$ is the
	anti-symmetric vorticity tensor, $\sigma^{\alpha\beta}\equiv \nabla^{\langle\alpha} u^{\beta\rangle}
	=\frac{1}{2}\left(\nabla^{\alpha} u^{\beta}+\nabla^{\beta} u^{\alpha}\right)-\frac{1}{3}
	\theta \Delta^{\alpha \beta}$ is the symmetric-traceless tensor and
	$\theta \equiv \partial_{\mu} u^{\mu}$ is the expansion scalar.
	The fourth-rank projection tensor is defined as
	$\Delta^{\mu\nu}_{\alpha\beta}=\frac{1}{2}\left(\Delta^{\mu}_{\alpha}\Delta^{\nu}_{\beta}+\Delta^{\mu}_{\beta}\Delta^{\nu}_{\alpha}\right)-\frac{1}{3}\Delta^{\mu\nu}\Delta_{\alpha\beta}$.
	
	
The manuscript is organized as follows:~In Sec.~\ref{sec:level2} we discuss the equation of motion for the
	relativistic magnetohydrodynamics, the corresponding energy-momentum tensors and the definition of some quantities
	related to the kinetic theory description used in the next section. In Sec.~\ref{sec:formalism} we derive the first
	and second order corrections to the single particle distribution function and the corresponding dissipative
	fluxes from the Boltzmann equation. Here we also discuss the Navier-Stokes limit and the Wiedemann-Franz law.
	Finally we conclude and summarise our study in Sec.~\ref{sec:conc}.
	
	\section{\label{sec:level2}Relativistic magnetohydrodynamics equations}
	\subsection{Equations of motion }
	Here we discuss the essential fluid equations briefly. The magnetohydrodynamics equations are consists of
	energy-momentum conservation equations for fluid and electromagnetic field and the Maxwell's equations.
	These set of conservation equations are closed with an Equation of State (EoS) relating fluid pressure, energy, and
	number density and a constitutive equation for the charge four current (see Ref.~\cite{Panda:2020zhr} for details).
	In presence of the electromagnetic field, there exists an external force on charged fluid, and the energy-momentum conservation
	takes the following form
	\begin{equation}\label{eq:partialTmunu}
		\partial_{\mu}T^{\mu\nu}_{f}=F^{\nu\lambda}J_{\lambda}.
	\end{equation}
	Here
	\begin{equation}\label{eq:emtensor}
		F^{\mu\nu}=E^{\mu}u^{\nu}-E^{\nu}u^{\mu}+\epsilon^{\mu\nu\alpha\beta}u_{\alpha}B_{\beta},
	\end{equation}
	and its dual given by
	\begin{equation}\label{eq:dual}
		\tilde{F}^{\mu\nu}=B^{\mu}u^{\nu}-B^{\nu}u^{\mu}-\epsilon^{\mu\nu\alpha\beta}u_{\alpha}E_{\beta},
	\end{equation}
	where $E^{\mu}=F^{\mu\nu}u_{\nu}$, $B^{\mu}=\tilde{F}^{\mu\nu} u_{\nu}$, $J^{\lambda}$ is the four-charge current.
	We note that $E^{\mu}u_{\mu}=B^{\mu}u_{\mu}=0$. $F^{\mu\nu}$ obeys the  Maxwell's equations
	\begin{eqnarray}{\label{eq:maxwelleqn}}
		\partial_{\mu}F^{\mu\nu} &=& J^{\nu}, \\
		\partial_{\mu}\tilde{F}^{\mu\nu} &=& 0.
	\end{eqnarray}
Later we also need the energy-momentum tensor and particle current defined in terms of moments in the following form
	\begin{eqnarray}{\label{eq:TmunuKin}}
		T^{\mu\nu}_f &=& \int_{}^{}dp p^{\mu}p^{\nu} \left(f+\bar{f}\right), \\
		{\label{eq:NmuKin}}
		N^{\mu}_f &=& \int_{}^{}dp p^{\mu} \left(f-\bar{f}\right).
	\end{eqnarray}
where $dp = g d^3 \mathbf{p}/[(2\pi)^3 p^0]$ with $p^0 =\sqrt{\mathbf{p}^2 +m^2}$, $m$ being the rest mass, g is the degeneracy factor (spin degeneracy for this case has been considered to be 1) with $\bar{f}$ is the contribution from the anti-particles.
	The four charge current is given as $J_f^{\mu}=q N^{\mu}$ where $q$ is the electric charge. In Landau frame $N^{\mu}=n_f u^{\mu}+V_{f}^{\mu}$
	where the dissipative part $V_{f}^{\mu}= \Delta^{\mu\nu}N_{\nu}$ and the charge density $n=N^{\mu}u_{\mu}$.
	

	The single particle equilibrium distribution function in Eq.~\eqref{eq:TmunuKin} and Eq.~\eqref{eq:NmuKin} is given by
	\begin{equation}\label{eq:distfunc}
		f_0=\frac{1}{e^{\beta(u \cdot p)-\alpha}+r},
	\end{equation}
	here $\beta=T^{-1}$ where $T$ is the temperature, $u^{\mu}$ is the fluid four-velocity, $p^{\mu}$ is the four-momentum, $\alpha=\mu\beta$ , and  $\mu$ is the chemical potential, $r=\pm 1,0$ for the fermions, bosons and for the Boltzmann case respectively, also $\alpha\rightarrow -\alpha$ for anti-particles.

	The energy-momentum tensor for the electromagnetic field  is
	\begin{eqnarray}\label{eq:emtensorforB}\nonumber
		T^{\mu\nu}_{EM}&=&\left(\frac{B^2+E^{2}}{2}\right)u^{\mu}u^{\nu}-\left(\frac{B^{2}+E^2}{2}\right)\Delta^{\mu\nu}-B^2 b^{\mu}b^{\nu} \\
		&&-E^{2}e^{\mu}e^{\nu}+2Q^{(\mu}u^{\nu)},
	\end{eqnarray}
	where $B^{\mu}=Bb^{\mu}$, $E^{\mu}=Ee^{\mu}$ ,  $b^{\mu}b_{\mu}=-1$, $e^{\mu}e_{\mu}=-1$, $Q^{\mu}=\mathcal{E}^{\mu\lambda\rho}E_{\lambda}B_{\rho}$ with $\mathcal{E}^{\mu\lambda\rho}=\epsilon^{\mu\lambda\rho\tau}u_{\tau}$ and $b^{\mu}u_{\mu}=e^{\mu}u_{\mu}=0$. We also define the second rank antisymmetric tensor  $B^{\mu\nu}=\epsilon^{\mu\nu\alpha\beta}u_{\alpha}B_{\beta}=-Bb^{\mu\nu}$ where $B^{\mu\nu}B_{\mu\nu}=2B^{2}$ and the corresponding normalised tensor  $b^{\mu\nu}=-\frac{B^{\mu\nu}}{B}$
	with the properties that $b^{\mu\nu}u_{\nu}=b^{\mu\nu}b_{\nu}=0$ and $b^{\mu\nu}b_{\mu\nu}=2$.
	We can now write the total $T^{\mu\nu}$ as $T^{\mu\nu}=T^{\mu\nu}_{EM}+T^{\mu\nu}_f$.
	For the non-dissipative fluid $T^{\mu\nu}_f=\epsilon u^{\mu}u^{\nu}-P \Delta^{\mu\nu}$.
	Hence the energy-momentum tensor for the non-dissipative fluid in presence of electromagnetic field takes the following form:
	\begin{eqnarray}\label{eq:idealmhd}\nonumber
		T^{\mu\nu}_{tot(0)}&=&\left(\epsilon+\frac{B^2+E^{2}}{2}\right)u^{\mu}u^{\nu}-\left(P+\frac{B^2+E^2}{2}\right)\Delta^{\mu\nu}\\
		&& -B^2 b^{\mu}b^{\nu}-E^{2}e^{\mu}e^{\nu}+2Q^{(\mu}u^{\nu)}.
	\end{eqnarray}
	Whereas for the dissipative fluid in EM field we have:
	\begin{eqnarray}{\label{eq:totalTmunu}}\nonumber
		T^{\mu\nu}_{tot}&=&\left(\epsilon+\frac{B^2+E^{2}}{2}\right)u^{\mu}u^{\nu}-\left(P+\Pi+\frac{B^2+E^{2}}{2}\right)\Delta^{\mu\nu}\\
		&&+\pi^{\mu\nu} -B^2b^{\mu}b^{\nu}-E^{2}e^{\mu}e^{\nu}+2Q^{(\mu}u^{\nu)}.
	\end{eqnarray}
	Later on in our calculation, we need the following expressions, which are obtained from the energy-momentum conservation equation
	and the thermodynamic integrals given in Ref.~\cite{Panda:2020zhr}.

\begin{widetext}
\begin{eqnarray}
\dot{\alpha}&=&\frac{1}{D_{20}}\left[J_{20}^{(0)-}\theta\left(\epsilon +P +\Pi\right)-J_{30}^{(0)+}\left(n_f \theta +\partial_{\mu}V_f^{\mu} \right) +J_{20}^{(0)-}\left(-\pi^{\mu\nu}\sigma_{\mu\nu}+qE^{\mu}V_{f\mu}\right)\right],
\nonumber
\\
\dot{\beta}&=&\frac{1}{D_{20}}\left[J_{10}^{(0)+}\theta\left(\epsilon +P +\Pi\right) -J_{20}^{(0)-}\left(n_f \theta +\partial_{\mu}V_f^{\mu} \right) +J_{10}^{(0)+}\left(-\pi^{\mu\nu}\sigma_{\mu\nu}+qE^{\mu}V_{f\mu} \right) \right],
\nonumber
\\
\dot{u}^{\mu}&=&\frac{1}{\epsilon+P}\left[\frac{n_f}{\beta}\left(\nabla^{\mu}\alpha-h \nabla^{\mu}\beta\right)-\Pi \dot{u}^{\mu}+\nabla^{\mu}\Pi-\Delta^{\mu}_{\nu}\partial_{\rho}\pi^{\rho \nu}\right] +\frac{1}{\epsilon+P}\left[q n_f E^{\mu}-qB b^{\mu\nu}V_{f \nu} \right],
\end{eqnarray}
\end{widetext}	
where $D_{20}=J^{(0)+}_{30}J^{(0)+}_{10}-J_{20}^{(0)-}J_{20}^{(0)-}$, $h=\frac{\epsilon+P}{n_f}$ and $\sigma^{\mu\nu}=\Delta^{\mu\nu}_{\alpha\beta}\nabla^{\alpha}u^{\beta}$.
	\section{FORMALISM AND BOLTZMANN EQUATION}
	\label{sec:formalism}
	\subsection{Boltzmann Equation}
	The relativistic Boltzmann equation(RBE) is given by
	\begin{equation}\label{RBE}
		p^{\mu}\partial_{\mu}f +qF^{\mu\nu}p_{\nu}\frac{\partial}{\partial p^{\mu}}f= C[f],
	\end{equation}
	where $f$ is the distribution function,  $q$ is the electric charge and $C[f]$ is the collision kernel.
	Here we take the collision kernel as $C[f]=-\frac{u\cdot p}{\tau_c}\delta f$ where $\tau_c$ is the relaxation time
	and $\delta f=f-f_{0}$ is the deviation from the local-equilibrium distribution function $f_{0}$.
	For this collision kernel we get
	\begin{equation}\label{eq:modrbe}
		p^{\mu}\partial_{\mu}f +qF^{\mu\nu}p_{\nu}\frac{\partial}{\partial p^{\mu}}f= -\frac{u\cdot p}{\tau_c}\delta f.
	\end{equation}
	\subsection{First order and second order derivation}
	Here we use the techniques similar to Ref.~\cite{Panda:2020zhr} in order to calculate $\delta f$ corrections. Eq.~\eqref{eq:modrbe} can be written as a power series expansion of the following form
	\begin{equation}\label{eq:fexpansion}
		f=\sum_{n=0}^{\infty}\left(-1\right)^n \left(\frac{\tau_c}{u.p}\right)^n\left(p^{\mu}\partial_{\mu}+qF^{\mu\nu}p_{\nu}\frac{\partial}{\partial p^{\mu}}\right)^n f_0.
	\end{equation}
	In the non-resistive case, one had two expansion parameter, viz. $\mathrm{Kn} = \tau_c \partial_{\mu}$ and $\chi = qB\tau_c /T$. In that case electric field was not an independent degree of freedom but was related to the magnetic field through the relation $\mathbf{E}=-\mathbf{v\times B}$. However, on lifting this assumption of infinite conductivity and including the effect of electric field explicitly in our calculation, we see that there is yet another expansion parameter $\xi=qE\tau_c/T$ apart from $\mathrm{Kn}$ and $\chi$.
	
	Now truncating it upto second order we get
	\begin{eqnarray}\label{eq:uptosecondorderf}
		f=f_0+\delta{f}^{(1)}+\delta{f}^{(2)},
	\end{eqnarray}
	where \begin{eqnarray}{\label{eq:deltaf1}}
		\delta{f}^{(1)}&=&-\frac{\tau_c}{u\cdot p}\left(p^{\mu}\partial_{\mu}f_0+\beta qE^{\nu}p_{\nu}f_0\tilde{f_0} \right),\\
		\delta{f}^{(2)}&=&\mathcal{F}_{1}+\mathcal{F}_{2}+\mathcal{F}_{3}+\mathcal{F}_{4}, \\ \nonumber
		\mathcal{F}_{1}&=&\frac{\tau_c}{u\cdot p}p^{\mu}\partial_{\mu}\left(\frac{\tau_c}{u\cdot p}p^{\sigma}\partial_{\sigma}f_0\right),\\ \nonumber
		\mathcal{F}_{2}&=&\frac{\tau_c}{u \cdot p}p^{\mu}\partial_{\mu}\left(\frac{q\tau_c}{u\cdot p}f_0\tilde{f}_0\beta (E\cdot p)\right),\\ \nonumber
		\mathcal{F}_{3}&=&\frac{q\tau_c}{u.p}F^{\mu\nu}p_{\nu}\frac{\partial}{\partial p^{\mu}}\left(\frac{\tau_c}{u\cdot p}p^{\sigma}\partial_{\sigma}f_0\right),\\ \nonumber
		\mathcal{F}_{4}&=&\frac{q^2\tau_c}{u\cdot p}F^{\mu\nu}p_{\nu}\frac{\partial}{\partial p^{\mu}}\left(\frac{\tau_c}{u\cdot p}f_0\tilde{f}_0 \beta (E\cdot p)\right).
	\end{eqnarray}
	Similarly the correction for anti-particles (${\delta \bar{f}}$) are calculated.
	\subsection{First order evolution equations}
	
	\subsubsection{Bulk Stress}
	We use Eq.~\eqref{eq:deltaf1} to calculate  the first-order dissipative fluxes.
	The bulk stress for the first order is given by
	\begin{equation}\label{eq:firstbulk}
		\Pi_{(1)}=-\frac{\Delta_{\mu\nu}}{3}\int_{}^{}dp p^{\mu}p^{\nu}\left(\delta f^{(1)} + \delta\bar{ f}^{(1)}\right).
	\end{equation}
	After some calculation we get 
	\begin{equation}\label{eq:resultbulk1}
		\Pi_{(1)}=-\tau_c \beta_{\Pi}\theta,
	\end{equation}
	where  $\beta_{\Pi}=\frac{5\beta}{3}J_{42}^{(1)+} +  \mathcal{X} J_{31}^{(0)+}-\mathcal{Y} J_{21}^{(0)-}$ with $$\mathcal{X}=\frac{J_{10}^{(0)+}\left( \epsilon+P\right)-J_{20}^{(0)-}n_f}{D_{20}},$$ and $$\mathcal{Y}=\frac{J_{20}^{(0)-}\left( \epsilon+P\right)-J_{30}^{(0)+}n_f}{D_{20}}.$$
	
	\subsubsection{Diffusion Current}
	The diffusion current for the first order is given by
	\begin{equation}\label{eq:firstdiffusion}
		V^{\mu}_{(1)}=\Delta^{\mu}_{\alpha}\int_{}^{}dp p^{\alpha}\left(\delta f^{(1)} - \delta\bar{ f}^{(1)}\right).
	\end{equation}
	Using Eq.~\eqref{eq:deltaf1} we get 
	\begin{equation}\label{eq:resultdiffusion1}
		V^{\mu}_{(1)}=\tau_c \beta_V \left(\nabla^{\mu}\alpha+ \beta  q E^{\mu}\right),
	\end{equation}
	where $\beta_V =\frac{1}{h}J_{21}^{(0)-}-J_{21}^{(1)-}$.
	
	\subsubsection{Shear Stress}
The shear stress for the first order is given by
	\begin{equation}\label{eq:firstshear}
		\pi^{\mu\nu}_{(1)}=\Delta^{\mu\nu}_{\alpha\beta}\int_{}^{}dp p^{\alpha}p^{\beta}\left(\delta f^{(1)} + \delta\bar{ f}^{(1)}\right),
	\end{equation}
	Using Eq.\eqref{eq:deltaf1} we get 
	\begin{equation}\label{eq:resultshear1}
		\pi^{\mu\nu}_{(1)}=2\tau_c \beta_{\pi}\sigma^{\mu\nu},
	\end{equation}
	where $\beta_{\pi}=\beta J_{42}^{(1)+}$.
	
	From the above discussion we see that the first order viscous terms are independent of
	the electromagnetic field but the diffusion current has contributions from the $E^{\mu}$.
	
	\subsection{Second order evolution equations}
	Here we evaluate the second order equations for dissipative stresses.

	\subsubsection{Bulk Stress}
	Similar to the first-order case, the expression for the second order bulk stress is given by
		\begin{eqnarray}\label{eq:bulk2}
		\Pi_{(2)}&=&-\frac{\Delta_{\alpha\beta}}{3}\int dp p^{\alpha}p^{\beta}\left(\delta f^{(2)}+\delta \bar{f}^{(2)}\right).
	\end{eqnarray}
	For this case the total bulk stress is composed of first and second order terms
	\begin{equation}\label{eq:evolutionbulk}
		\Pi=\Pi_{(1)}+\Pi_{(2)}.
	\end{equation}
	We obtain the evolution equation for the bulk stress from Eq.~\eqref{eq:bulk2} and Eq.~\eqref{eq:evolutionbulk}
	(see appendix.~\ref{app:bulk} for details)
\begin{widetext}
\begin{eqnarray}
\frac{\Pi}{\tau_c}&=&-\dot{\Pi}-\delta_{\Pi\Pi}\Pi \theta +\lambda_{\Pi\pi}\pi^{\mu\nu}
		\sigma_{\mu\nu}-\tau_{\Pi V}V\cdot \dot{u}-\lambda_{\Pi V}V\cdot \nabla \alpha -l_{\Pi V}\partial \cdot V-\beta_{\Pi}\theta-
		 qB\lambda_{\Pi VB} b^{\mu\beta}V_{\beta}V_{\mu}
		\nonumber
		\\
		&&+\tau_c \tau_{\Pi V B}\dot{u}_{\alpha} qBb^{\alpha\beta}V_{\beta}- q
		\delta_{\Pi V B}\nabla_{\mu}\left(\tau_c B b^{\mu\beta}V_{\beta} \right)  -q^2\tau_{c}\chi_{\Pi EE}E^{\mu}E_{\mu}.
		\label{eq:2bulkevolutionexp}
\end{eqnarray}
\end{widetext}
	Where the  transport coefficients appearing in Eq.~\eqref{eq:2bulkevolutionexp} are listed in Table~\ref{table:bulkjmn} and rest of the coefficients have usual meaning as for the ideal MHD case Ref.~\cite{Panda:2020zhr}.

	\subsubsection{Diffusion Current}	
	The expression for the diffusion current is obtained in a similar manner with the exception that now the particle and antiparticle contribution is not additive as per the definition 
	\begin{eqnarray}\label{eq:2Diffusion}
		V^{\mu}_{(2)}&=&\Delta^{\mu}_{\alpha}\int dp p^{\alpha} \left(\delta f^{(2)}-\delta \bar{f}^{(2)} \right),
	\end{eqnarray}
	\begin{equation}\label{eq:evolutiondiffusion}
		V^{\mu}=V^{\mu}_{(1)}+V^{\mu}_{(2)} .
	\end{equation}
	The second-order evolution equation for the diffusion current is obtained from Eq.~\eqref{eq:2Diffusion} and Eq.~\eqref{eq:evolutiondiffusion} (see appendix.~\ref{app:diffusion} for details)
	\begin{widetext}
	\begin{eqnarray}
	\frac{V^{\mu}}{\tau_c}&=&-\dot{V}^{\langle\mu\rangle}-V_{\nu}\omega^{\nu \mu}
		+\lambda_{VV}V^{\nu}\sigma^{\mu}_{\nu}-\delta_{VV}V^{\mu} \theta
		+\lambda_{V\Pi}\Pi\nabla^{\mu} \alpha -\lambda_{V\pi}\pi^{\mu \nu}\nabla_{\nu}\alpha
		-\tau_{V\pi}\pi^{\mu}_{\nu}\dot{u^{\nu}}-q B \delta_{V B} b^{\mu\gamma}V_{\gamma} \nonumber
		\\
		&& +\tau_{V\Pi}\Pi \dot{u^{\mu}}+l_{V\pi}\Delta^{\mu \nu}\partial_{\gamma}\pi^{\gamma}_{\nu}
		-l_{V\Pi}\nabla^{\mu}\Pi+\beta_V \nabla^{\mu} \alpha +\tau_c q B l_{V\pi B} b^{\sigma \mu}\partial^{\kappa}\pi_{\kappa\sigma}-q \tau_c \lambda_{VVB}B
		b^{\gamma \nu}V_{\nu}\sigma^{\mu}_{\gamma} \nonumber \\
		&& +\tau_c q B \tau_{V \Pi B}b^{\gamma\mu}\Pi \dot{u}_{\gamma} -\tau_c qB l_{V \Pi B} b^{\gamma\mu}
		\nabla_{\gamma}\Pi-q\tau_c \delta_{VVB} B b^{\mu\nu}V_{\nu}\theta -q \tau_c \mathbf{\rho}_{VVB} B b^{\gamma \nu}V_{\nu}
		\omega^{\mu}_{\gamma} 
	  \nonumber \\
		&&+ \chi_{VE}  q E^{\mu}+q\Delta^{\mu}_{\alpha}\chi_{VE}D\left(\tau_c E^{\alpha}\right)-q \tau_c \rho_{VE} E^{\mu}\theta	- q \tau_{VVB}\Delta^{\mu}_{\gamma}D\left(\tau_cBb^{\gamma \nu}V_{\nu} \right).
		\label{eq:fulldiffusion}
	\end{eqnarray}
	\end{widetext}

	The coefficients appearing for the resistive case are listed in Table~\ref{table:diffusionjmn}, the rest of the 
	transport coefficients are same as the ideal MHD case given in Ref.~\cite{Panda:2020zhr}.	
	\subsubsection{ Shear Stress }
	The expression for the second-order shear stress is obtained from the following definition
	\begin{eqnarray}\label{eq:shear2}
		\pi^{\mu\nu}_{(2)}&=&\Delta^{\mu\nu}_{\alpha\beta}\int dp p^{\alpha}p^{\beta} \left(\delta f^{(2)}+\delta \bar{f}^{(2)}\right).
	\end{eqnarray}
	Note that the total shear stress is the combination of first and second order terms	
	\begin{equation}
		\label{eq:2evolutionstress}
		\pi^{\mu\nu}=\pi^{\mu\nu}_{(1)} +\pi^{\mu\nu}_{(2)}.
	\end{equation}
Evaluating the integral in Eq.~\eqref{eq:shear2} (see Appendix~(\ref{app:shear}) for details) and adding
	it to Eq.~\eqref{eq:2evolutionstress} we get the evolution equation for the shear stress 
\begin{widetext}
\begin{eqnarray}
\frac{\pi^{\mu\nu}}{\tau_c}&=&-\dot{\pi}^{\left<\mu\nu\right>}+2\beta_{\pi}\sigma^{\mu\nu}
		+2\pi^{\langle\mu}_{\gamma}\omega^{\nu\rangle\gamma}-\tau_{\pi\pi}
		\pi^{\langle\mu}_{\gamma}\sigma^{\nu\rangle\gamma} -\delta_{\pi\pi}\pi^{\mu\nu}\theta  +\lambda_{\pi\Pi}\Pi\sigma^{\mu\nu}-\tau_{\pi V}V^{\langle\mu}\dot{u}^{\nu\rangle}-\tau_c qB\tau_{\pi VB} \dot{u}^{\langle\mu}b^{\nu\rangle\sigma} V_{\sigma}
		\nonumber \\
		&&+\lambda_{\pi V}V^{\langle\mu}\nabla^{\nu \rangle}\alpha -l_{\pi V}\nabla^{\langle\mu}
		V^{\nu\rangle} +\delta_{\pi B}\Delta^{\mu\nu}_{\eta \beta}q B b^{\gamma \eta}g^{\beta \rho}
		\pi_{\gamma\rho} - qB\lambda_{\pi VB} V_{\gamma}b^{\gamma\langle\mu}V^{\nu\rangle}
		-q \delta_{\pi VB}   \nabla^{\langle\mu}\left(\tau_c B^{\nu\rangle\gamma}V_{\gamma} \right) \nonumber
		\\
		&&+ q^2\tau_{c} \chi_{\pi EE} \Delta^{\mu\nu}_{\sigma \rho } E^{\sigma}E^{\rho}. 
		\label{eq:2evolutionexp}
\end{eqnarray}	
\end{widetext}

	The  coefficients in Eq.\eqref{eq:2evolutionexp} that appear for the resistive case only are listed in Table~\ref{table:shearjmn} , the rest of the coefficients are same as ideal MHD Ref.~\cite{Panda:2020zhr}.
	
\onecolumngrid

\begin{table}[h!]
\begin{minipage}{\linewidth}
\centering
\begin{minipage}{0.4\linewidth}
 \centering
 \begin{tabular}{|l|l|l|}
 \hline
 Transport	& CE	&	Denicol	\\
Coefficients & 	&	et al.	\\ \hline
$\tau_{\Pi V}$ 	& 	$  0  $ 	& 	$0$  \\ \hline
$\chi_{\Pi EE}$ & 	${\beta^2 P}/{36}$ &	$-$ \\ \hline
$\lambda_{\Pi V}$ 	& 	${1}/{(3\beta)}$ & $ 0$ \\ \hline
$l_{\Pi V}$ 	& 	$0$ &	$0$	  \\ \hline
$\lambda_{\Pi V B}$ 	    & 	$3/(\beta P)$      & $-$ 	\\ \hline
 \end{tabular}\\[0.9em]
 (a)
 \end{minipage}
\begin{minipage}{0.5\linewidth}
 \centering
 \begin{tabular}
{|l|l|l|}
 \hline
Transport	& CE	&	Denicol	\\
Coefficients & 	& et al.		\\ \hline
$\lambda_{VV}$ 	& 	$2/5$ &	$3/5$ \\ \hline
$\delta_{VV}$ 	& 	$22/3$ &  $1$	 \\ \hline
$\delta_{VB}$ 	& 	$2\beta$  & ${5\beta}/{12}$	  \\ \hline
$\rho_{V E}$ 	& 	$P\beta^2/18$ & $-$   \\ \hline
$\chi_{V E}$ 	& 	${P \beta^2}/{12}$ & ${P \beta^2}/{12}$   \\ \hline
 \end{tabular}\\[0.9em]
 (b)
\end{minipage}
\begin{minipage}{0.4\linewidth}
 \centering
 \begin{tabular}
{|l|l|l|}
 \hline
Transport	& CE	&	Denicol	\\
Coefficients & 	& et al.		\\ \hline
$\tau_{\pi V}$ 	& 	$12/(5\beta)$ 	&	$0$ 		\\ \hline
$l_{\pi V}$ & 	$12/(5\beta)$  &	$0$		\\ \hline
$\lambda_{\pi V}$ 	& 	$11/5\beta$ 	&	$0$	\\ \hline
$\lambda_{\pi V B}$ 	& 	$24/5{(\beta P)}^{-1}$ 	&	$-$	\\ \hline
$\chi_{\pi E E}$ 	& 	$2\beta^2 P/15 $  &	$-$	\\ \hline
 \end{tabular}\\[0.9em]
 (c)
\end{minipage}
\end{minipage}
\caption{\label{Table:1}(a)  Transport coefficients for the bulk-stress
for a massless Boltzmann gas~(result for particles only and for zero chemical potential) calculated using CE method (this work) and compared with the results from the moment method ref.~\cite{Denicol:2019iyh}
(b) same as Table \ref{Table:1}(a) but for diffusion current (c) same as Table \ref{Table:1}(a) but for
shear stress.}
\end{table}

\twocolumngrid

	\subsection{Navier-stokes equations}
	Here we keep the terms which are only first-order in gradient in Eq.\eqref{eq:2bulkevolutionexp}, Eq.\eqref{eq:fulldiffusion}, Eq.\eqref{eq:2evolutionexp} and get the Navier-Stokes limit:
	\begin{eqnarray}
		\label{eq: bulk_evolution3}
		&&\frac{\Pi}{\tau_c}=-\beta_{\Pi}\theta,\\
		\label{eq:diffusionEvolution4}
		&&V^{\mu}+qB\tau_c \delta_{VB}b^{\mu\nu}V_{\nu}-q\tau_c \beta \beta_{V}E^{\mu}=\tau_c \beta_{V}\nabla^{\mu}\alpha ,\\
		\label{eq:shear_evolution4}
		&&\left(\frac{g^{\mu\gamma}g^{\nu\rho}}{\tau_c}-\delta_{\pi B}\Delta^{\mu\nu}_{\eta \beta}q B b^{\gamma \eta}g^{\beta \rho}\right)\pi_{\gamma\rho}=2\beta_{\pi}\sigma^{\mu\nu}.
	\end{eqnarray}
In the power counting scheme, the electric field $E^{\mu}$ is considered $\mathcal{O}(\partial)$ as given in Ref.~\cite{Hernandez:2017mch}. Using the same projection operators as used in Ref.~\cite{Panda:2020zhr} we get the coefficients for the shear, bulk, and diffusion. It turned out that the first-order transport coefficients for the shear
and the bulk viscosity is the same as the ideal MHD case \cite{Panda:2020zhr}, however, for diffusion, we get new transport coefficients. The decomposition of the diffusion four current in terms of the projector expressed as 
	\begin{eqnarray}\nonumber
		V_{\nu}&=&\left(\kappa_{\parallel}P^{\parallel}_{\delta\nu}+\kappa_{\bot}P^{\bot}_{\delta\nu}
		+\kappa_{\times}P^{\times}_{\delta\nu}\right)\partial^{\delta}\alpha \\ 
		&&+\frac{1}{q}\left(\sigma_{\parallel}P^{\parallel}_{\delta\nu}+\sigma_{\bot}P^{\bot}_{\delta\nu}
		+\sigma_{\times}P^{\times}_{\delta\nu}\right)E^{\delta}.
	\end{eqnarray}
	Here we have used the Ohmic law for current in a conducting fluid
	\begin{eqnarray}
		J_{ind}^{\mu}=\sigma_{E}^{\mu\nu}E_{\nu},
	\end{eqnarray}
	where $\sigma_E^{\mu\nu}$ is the electrical conductivity tensor.

	Putting the above equation in Eq.~\eqref{eq:diffusionEvolution4}; using the properties of the projection operators and comparing both sides for the $\partial^{\delta}\alpha$ we get
	\begin{eqnarray}\nonumber
		\kappa_{\parallel}&=&\tau_c \beta_{V},\\ \nonumber
		\kappa_{\bot}&=&\frac{\tau_c \beta_{V}}{1+\left(qB \tau_c \delta_{VB}\right)^2}, \\ 
		\kappa_{\times}&=&\kappa_{\bot} q B \tau_c \delta_{VB}.
	\end{eqnarray}

	
	Similarly comparing the coefficients of $E^{\delta}$ we get 
	
	\begin{eqnarray}
	\nonumber{\label{eq:conductivity}}
		\sigma_E^{\parallel}&=&q^2 \tau_c \beta \beta_{V} ,\\
		\nonumber
		\sigma_{E}^{\bot}&=&\frac{q^2 \tau_c \beta \beta_{V}}{1+\left(qB \tau_c \delta_{VB}\right)^2},\\
		\sigma_{E}^{\times}&=&\frac{q^3 B \tau_c^2 \beta \beta_{V}  \delta_{VB}}{1+\left(qB \tau_c \delta_{VB}\right)^2}.
	\end{eqnarray}

		
	The above relations are the kinetic version of the Wiedemann-Franz law which is $\sigma= q^2 \beta \kappa$. Moreover, if we set all the dissipative quantities zero, then we will get $\nabla^{\mu}\alpha=-q\beta E^{\mu}$. As expected, $\sigma_E^{\parallel}$ is independent of the magnetic field and proportional to $T^{2}$ ; $\sigma_{E}^{\bot}$, and $\sigma_{E}^{\times}$ decreases for increasing magnetic fields \cite{article:Kremer,Dash:2020vxk,Satapathy:2021cjp}. It may be helpful for application in the Cooper-Frye prescription if we express the 
	$\delta f$ correction in terms of the dissipative quantities. In that case, the $\delta f^{(1)}$ (Eq.\eqref{eq:deltaf1}) can be written as:
\begin{eqnarray}\nonumber
\delta f^{(1)}=\frac{f_{0}\Bar{f}_{0}\tau_c}{u\cdot p}\left(\mathcal{A} \Pi+\mathcal{B}^{\beta}V_{\beta}+\mathcal{C}^{\gamma\rho}\pi_{\gamma\rho} \right),
\end{eqnarray}
where
\begin{eqnarray}
\mathcal{A}&=&-\frac{1}{\tau_c \beta_{\Pi}}\Bigg[\frac{(u\cdot p)^2}{D_{20}}\left(J_{20}^{(0)-}\left(\epsilon+P\right)-J_{30}^{(0)+}n_f\right)
\nonumber
\\
&& -\frac{(u\cdot p)}{D_{20}}\left(J_{10}^{(0)+}\left(\epsilon+P\right)-J_{20}^{(0)-}n_f\right)+\frac{\beta \Delta_{\mu\beta}p^{\mu}p^{\beta}}{3}\Bigg],
\nonumber
\\
\mathcal{B}^{\beta}&=&-\frac{p^{\beta}}{\tau_c \beta_{V}}+\frac{n_f(u\cdot p)p^{\beta}}{\tau_c \beta_{V}\left(\epsilon +P\right)}-\left(1-\frac{(u\cdot p)}{h}\right)\frac{qB p^{\mu}\delta_{VB}b^{\beta}_{\mu}}{\beta_{V}},
\nonumber
\\ \nonumber
\mathcal{C}^{\gamma\rho}&=&\frac{\beta p^{\beta}p^{\kappa}}{2 \beta_{\pi}}\left(\frac{g^{\gamma}_{\kappa}g^{\rho}_{\beta}}{\tau_c}-\delta_{\pi B}\Delta^{\mu\nu}_{\eta \tau}q B b^{\gamma\eta}g^{\tau \rho}g_{\beta\nu}g_{\kappa\mu}\right).
\end{eqnarray}
Here we tried to follow the custom to express $\delta f $ in powers of 
$p^{\mu}$. It is interesting to note that we cannot do that for $\mathcal{B}^{\beta}$ and $\mathcal{C}^{\gamma\rho}$, where the correction due to the magnetic fields appears in the first-order.

	\section{Conclusion}	
	\label{sec:conc}
	In this work, we derive the second-order relativistic resistive dissipative magnetohydrodynamics equations using the relaxation time approximation of the collision kernel in the relativistic Boltzmann equation. The resistive MHD implies a non-zero electric field inside the fluid; we found novel transport coefficients originating due to the coupling of the electromagnetic field to the usual dissipative forces. We also calculate the values of these new transport coefficients for a Boltzmann gas in the massless limit and compare it to the fourteen-moment method~\cite{Denicol:2019iyh}. We derive the Navier-Stokes limit of the second-order equations using power counting, and subsequently, we show the relationship between particle diffusion and electrical conductivity, a.k.a. Wiedemann-Franz law.
	
	We wish to further extend the current formulation to curved space-time, which is relevant for cosmological problems \cite{Chabanov:2021dee}.
	In the future, one can study the evolution of QGP by numerically solving the equations obtained here for the resistive case and estimate possible uncertainties due to the magnetic field in the extracted values of QGP transport coefficients. Finally, it will be
	interesting to work in the quantum regime and study the effect of spin degrees of freedom of the quasiparticles. The study of
	spin-magnetohydrodynamics~\cite{Israel:1978up,Brodin:2006be,Singh:2021man} may enable us to understand the phenomena of polarisation of vector mesons produced in high-energy heavy-ion collisions.

\begin{acknowledgments}
AP acknowledges the CSIR-HRDG financial support. AD and VR acknowledge support from the DAE, Govt. of India. RB and VR acknowledge financial support from the DST Inspire faculty research grant (IFA-16-PH-167), India.
\end{acknowledgments}

\onecolumngrid

\appendix	
		\section{Second order relaxation equations}
		Here we give the detailed calculations of the second order dissipative quantities. The explicit dependence of the anti-particles is not shown in the calculation, but they appear in our final results.
			
		\subsection{Bulk stress}{\label{app:bulk}}
		Let us consider the bulk viscous case first. From Eq.~\eqref{eq:bulk2} we get:
		
		\begin{eqnarray}
			\Pi&=&-\frac{\Delta_{\alpha\beta}}{3}\int_{}^{}dp p^{\alpha}p^{\beta}\left(\delta f^{(2)} + \delta \bar{f}^{(2)}\right),\\
			\Pi&=&\mathcal{I}_{1}+\mathcal{I}_{2}+\mathcal{I}_{3}+\mathcal{I}_{4}.
		\end{eqnarray}
		
		For convenience we split $\mathcal{I}_{1}$ into three terms as follows:
		
		\begin{eqnarray}
			\mathcal{I}_{1}=\mathcal{A}+ \mathcal{B}+\mathcal{C},
		\end{eqnarray}
		
		where
		\begin{eqnarray}
			\nonumber
			\mathcal{A}&=&-\frac{\Delta_{\alpha\beta}}{3}\int_{}^{}dp p^{\alpha}p^{\beta}\tau_c D\left[\frac{\tau_c}{u \cdot p}p^{\rho}\partial_{\rho} f_0\right],\\
			\nonumber
			\mathcal{B}&=&-\frac{\Delta_{\alpha\beta}}{3}\int_{}^{}dp p^{\alpha}p^{\beta}\frac{\tau_c}{u \cdot p}p^{\mu}\nabla_{\mu}\left(\tau_c \dot{f_0}\right),\\
			\nonumber
			\mathcal{C}&=&-\frac{\Delta_{\alpha\beta}}{3}\int_{}^{}dp p^{\alpha}p^{\beta}\frac{\tau_c}{u \cdot p}p^{\mu}\nabla_{\mu}\left(\frac{\tau_c p^{\rho}}{u \cdot p}\nabla_{\rho}f_0\right).
		\end{eqnarray}
We  carry out these integrals one-by-one and we get:
\begin{eqnarray}
\mathcal{A}&=&-\tau_c \dot{\Pi}+\frac{2\tau_c^2}{3h}J_{31}^{(0)+}\dot{u}_{\alpha}\nabla^{\alpha}\alpha -\frac{2\tau_c^2}{3}J_{21}^{(0)-}\dot{u}_{\alpha}\nabla^{\alpha}\alpha -\frac{2\tau_c^2\beta}{3h}J_{31}^{(0)+}\dot{u}_{\alpha}qBb^{\alpha\beta}V_{\beta}
+\frac{2\tau_c^2\beta}{3(\epsilon+P)}J_{31}^{(0)+}\dot{u}_{\alpha}qE^{\alpha},
\nonumber
\\
\mathcal{B}&=&\frac{5\tau_c^2}{3}\nabla_{\mu}\left(\beta J_{42}^{(1)+}\dot{u}^{\mu}\right)+\frac{5\tau_c^2}{3}\theta\left[\left(J_{31}^{(0)+}+J_{42}^{(1)+}\right)\dot{\beta}-\left(J_{31}^{(1)-}+J_{42}^{(2)-}\right)\dot{\alpha}\right],
\nonumber
\\
\mathcal{C}&=&\frac{5\tau_c^2 \beta }{9} \left(7J_{63}^{(3)+}+\frac{23}{3}J_{42}^{(1)+}\right)\theta^{2} +\frac{5\tau_c^2}{3}\nabla_{\mu}\left[\nabla^{\mu}\alpha\left(\frac{1}{h}J_{42}^{(1)-}-J_{42}^{(2)-}\right)\right] +\frac{\tau_c^2 \beta}{3}\left(7J_{63}^{(3)+}+J_{42}^{(1)+}\right)\sigma^{\mu\nu}\sigma_{\mu\nu} 
\nonumber
\\
&&+\frac{5\tau_c^2}{3}\nabla_{\mu}\left[-J_{42}^{(1)+}\beta\dot{u}^{\mu}-\frac{J_{42}^{(1)+}\beta qBb^{\mu \nu}V_{\nu}}{\epsilon+P}\right] +\frac{5\tau_c^2}{3}\nabla_{\mu}\left[\frac{J_{42}^{(1)-}\beta q E^{\mu}}{h}\right].
\end{eqnarray}

		For the rest three terms after some algebra we get:		
		\begin{eqnarray}\nonumber
			\mathcal{I}_{2}&=&-\frac{2}{3}\tau_c^2 q\beta E_{\mu}J_{31}^{(1)-}\dot{u}^{\mu}-\frac{5}{3}\tau_c^2 q \nabla_{\mu}(\beta J_{42}^{(2)-}E^{\mu}), \\ \nonumber
			\mathcal{I}_{3}&=&-q \tau_{c}^{2} \Big[\frac{1}{3h}\left( 5 J_{42}^{(2)-} + 2J^{(1)-}_{31} - 5 J_{42}^{(3)+} - 2J^{(2)+}_{31}\right)E^{\mu}\nabla_{\mu}\alpha  
 + \frac{q \beta}{3h } \left( 5 J_{42}^{(2)-} + 2J^{(1)-}_{31}\right)E^{\mu}E_{\mu} 
\nonumber
\\ \nonumber
&&+ \frac{q\beta}{3h n_{f}} \left( 5 J_{42}^{(2)-} + 2J^{(1)-}_{31}\right)E^{\mu} B_{\mu\nu}V^{\nu}_{f}\Big],
\\
		\mathcal{I}_{4}&=&\frac{q^2 \tau_c^2}{3}\left(5 \beta  J_{42}^{(3)+}+ 2 \beta J_{31}^{(2)+}\right)E_{\mu}E^{\mu}.
		\end{eqnarray}

		
		\subsection{Diffusion}{\label{app:diffusion}}
The second order diffusion current $V^{\mu}_{(2)}$ is given by Eq.~\eqref{eq:2Diffusion}:
\begin{eqnarray}
			V^{\mu}_{(2)}&=&\Delta^{\mu}_{\alpha}\int_{}^{}dp p^{\alpha}\left( \delta f ^{(2)}- \delta \bar{f} ^{(2)}\right) \nonumber ,\\
			V^{\mu}_{(2)}&=& \mathcal{I}_{1}+\mathcal{I}_{2}+\mathcal{I}_{3}+\mathcal{I}_{4}.
		\end{eqnarray}
Like the previous case we split $\mathcal{I}_{1}$ into three terms as follows:
\begin{eqnarray}
			\mathcal{I}_{1}=\mathcal{A}+ \mathcal{B}+\mathcal{C},
		\end{eqnarray}
		where
		\begin{eqnarray}
			\nonumber
			\mathcal{A}&=&\Delta^{\mu}_{\alpha}\int_{}^{}dp p^{\alpha}\tau_c D\left[\frac{\tau_c}{u\cdot p}p^{\rho}\partial_{\rho}f_0\right], \nonumber \\
			\mathcal{B}&=&\Delta^{\mu}_{\alpha}\int_{}^{}dp p^{\alpha} \frac{\tau_c}{u\cdot p}p^{\sigma}\nabla_{\sigma}\left(\tau_c \dot{f}_0\right), \nonumber \\
			\mathcal{C}&=&\Delta^{\mu}_{\alpha}\int_{}^{}dp p^{\alpha}\frac{\tau_c}{u\cdot p}p^{\sigma}\nabla_{\sigma}\left(\frac{\tau_c p^{\rho}}{u\cdot p}\nabla_{\rho}f_0\right). \nonumber
		\end{eqnarray}
		
		After integrating we have:		
		\begin{eqnarray}\nonumber
			\mathcal{A}&=&-\tau_c\dot{V}^{\left< \mu \right>}-\tau_c^2 \Delta^{\mu}_{\gamma}D\left[\frac{qn_fB b^{\gamma\nu}V_{\nu}}{\epsilon+P}\right]-\Delta^{\mu}_{\nu}D\left[q\beta \tau_c^2 E^{\nu}J_{21}^{(1)-}\right], \\
			\nonumber
			\mathcal{B}&=&-\tau_c^2\nabla^{\mu}\left(J_{21}^{(0)-}\dot{\beta}-J_{21}^{(1)+}\dot{\alpha}\right)-\frac{\tau_c^2 \beta \dot{u}^{\mu}\theta }{3}\left(4J_{21}^{(0)-}+5J_{42}^{(2)-}\right) -\tau_c^2 \beta J_{21}^{(0)-}\dot{u}_{\gamma}\omega^{\gamma\mu}-\tau_c^2 \beta \dot{u}_{\gamma}\sigma^{\gamma\mu}\left(J_{21}^{(0)-}+2J_{42}^{(2)-}\right),
			\nonumber
			\\ \nonumber
			\mathcal{C}&=&-\frac{4\tau_c^2}{3}\theta\left(\frac{1}{h}J_{21}^{(0)+}-J_{21}^{(1)+}\right)\nabla^{\mu}\alpha  +\frac{4\tau_c^2}{3}J_{21}^{(0)-}\beta\dot{u}^{\mu}\theta
			-\tau_c^2\left(\frac{1}{h}J_{21}^{(0)+}-J_{21}^{(1)+}\right)\sigma^{\mu}_{\gamma}\nabla^{\gamma}\alpha +\tau_c^2J_{21}^{(0)-}\beta\dot{u}^{\gamma}\sigma^{\mu}_{\gamma}
			\\
			\nonumber
			&&-\tau_c^2\left(\frac{1}
			{h}J_{21}^{(0)+}-J_{21}^{(1)+}\right)\omega^{\mu}_{\gamma} \nabla^{\gamma}\alpha +\tau_c^2J_{21}^{(0)-}\beta\dot{u}^{\gamma}\omega^{\mu}_{\gamma}
			+\tau_c J_{21}^{(0)-} \omega^{\mu}_{\gamma}
			\left[\frac{\beta qBb^{\gamma\nu}V_{\nu}}{\epsilon+P}\right]-2\tau_c^2\left(\frac{1}{h}J_{42}^{(2)+}-J_{42}^{(3)+}\right)\sigma^{\mu}_{\gamma}\nabla^{\gamma}\alpha
			\\
			&&
			 +2\tau_c^2J_{42}^{(2)-}\beta\dot{u}^{\gamma}\sigma^{\mu}_{\gamma}-\frac{5\tau_c^2}{3}\theta\left(\frac{1}{h}J_{42}^{(2)+}-J_{42}^{(3)+}\right)\nabla^{\mu}\alpha  +\frac{5\tau_c^2}{3}J_{42}^{(2)-}\beta\dot{u}^{\mu}
			\theta-2\tau_c^2\Delta^{\mu}_{\rho} \nabla_{\gamma}\left(\beta J_{42}^{(2)-}\sigma^{\rho\gamma}\right)-\frac{5\tau_c^2}{3}\nabla^{\mu}\left(\beta J_{42}^{(2)-}\theta\right)
			\nonumber\\
			&&
			+\frac{4\tau_c^{2}}{3}J_{21}^{(0)-} \theta\left[\frac{\beta qBb^{\mu\nu}V_{\nu}}{\epsilon+P}\right]
			+\tau_c^2 J_{21}^{(0)-} \sigma^{\mu}_{\gamma}\left[\frac{\beta qBb^{\gamma\nu}V_{\nu}}{\epsilon+P}\right]
			+2\tau_c^2 J_{42}^{(2)-} \sigma^{\mu}_{\gamma} \left[\frac{\beta qBb^{\gamma\nu}V_{\nu}}{\epsilon+P}\right]
			+\frac{5\tau_c^2}{3}J_{42}^{(2)-}\theta \left[\frac{\beta qBb^{\mu\nu}V_{\nu}}{\epsilon+P}\right] \nonumber \\
			&&-\tau_c^2\left(\frac{4}{3}J_{31}^{(1)+}+\frac{5}{3}J_{42}^{(2)+}\right)\theta \left(\frac{q\beta n_f E^{\mu}}{\epsilon+P}\right) -2\tau_c^2J_{42}^{(2)+}\sigma^{\mu}_{\gamma}\left(\frac{q\beta n_f E^{\gamma}}{\epsilon+P}\right) -\tau_c^2 J_{31}^{(1)+}\left(\omega^{\mu}_{\gamma}+\sigma^{\mu}_{\gamma}\right)\frac{q\beta n_f E^{\gamma}}{\epsilon+P}.
		\end{eqnarray}

The rest of the terms give:		
\begin{eqnarray}
			\nonumber
			\mathcal{I}_{2}&=&q\tau_c^2 D\left(\beta J_{21}^{(1)+}E^{\mu} \right)+q \tau_c^2 \beta J_{21}^{(1)+}E^{\alpha}u^{\mu}\dot{u}_{\alpha} +q \tau_c^2 \beta J_{31}^{(2)+}\left( E_{\nu}\left(\omega^{\nu\mu}+\sigma^{\nu\mu}+\frac{\Delta^{\nu\mu}}{3}\theta\right)+E^{\mu}\theta\right)  \\ \nonumber
			&&+q \tau_c^2 \beta J_{42}^{(3)+}\left(E^{\mu}\theta + 2 E_{\rho}\left(\sigma^{\mu \rho}+\frac{\Delta^{\mu\rho}}{3}\theta\right)\right),  \\ \nonumber
			\mathcal{I}_{3}&=&q \tau_c^2 \left(\beta J_{31}^{(2)+}B^{\mu \nu}\left(\frac{1}{\epsilon+P}\left[\frac{n_f}{\beta}\left(\nabla_{\nu}\alpha\right)-\Pi \dot{u}_{\nu}+\nabla_{\nu}\Pi-\Delta_{\nu\mu}\partial_{\rho}\pi^{\rho \mu}+q n_f E_{\nu}-qB b^{\mu}_{\nu}V_{f \mu} \right]\right)\right)  \\ \nonumber
			&&+q \tau_c^2 \left(\beta J_{31}^{(2)+}E^{\mu}\theta + J_{20}^{(1)+}E^{\mu}\dot{\beta} \right)+q \tau_c^2 \left( - J_{20}^{(2)+} E^{\mu}\dot{\alpha}-J_{21}^{(2)+}B^{\mu}_{\nu}\nabla^{\nu}\alpha \right)  \\ \nonumber
			&&+q \tau_c^2  \left( \beta E_{\nu} J_{42}^{(3)+}\left( 2 \sigma^{\mu\nu}+\frac{5}{3}\Delta^{\mu\nu}\theta\right) + E^{\mu}J_{31}^{(2)+} \dot{\beta}\right) -q \tau_c^2 \left( E^{\mu} J_{31}^{(3)+}\dot{\alpha} \right), \\ 
			\mathcal{I}_{4}&=&-\beta q^2 \tau_c^2 J_{21}^{(2)-}E^{\nu}B^{\mu}_{\nu}.
		\end{eqnarray}


		\subsection{Shear stress}{\label{app:shear}}
		The second order shear stress $\pi^{\mu\nu}_{(2)}$ is given by Eq.~\eqref{eq:shear2}:
		
		\begin{eqnarray}
			\nonumber
			\pi^{\mu\nu}_{(2)}&=&\Delta^{\mu\nu}_{\alpha\beta}\int_{}^{}dp p^{\alpha}p^{\beta} \left(\delta f^{(2)}+\delta \bar{f}^{(2)}\right) \nonumber \\
			&=&\mathcal{I}_{1}+\mathcal{I}_{2}+\mathcal{I}_{3}+\mathcal{I}_{4}.
		\end{eqnarray}

		Again the first term can be divided into three parts and is given by :		
		\begin{eqnarray}
			\mathcal{I}_{1}=\mathcal{A}+ \mathcal{B}+\mathcal{C},
		\end{eqnarray}
		where
		\begin{eqnarray}
			\nonumber
			\mathcal{A}&=&\Delta^{\mu\nu}_{\alpha \beta}\int_{}^{}dp p^{\alpha}p^{\beta} \tau_c D\left[\frac{\tau_c}{u \cdot p}p^{\sigma}\partial_{\sigma} f_0\right],\\
			\nonumber
			\mathcal{B}&=&\Delta^{\mu\nu}_{\alpha \beta}\int_{}^{}dp p^{\alpha}p^{\beta} \frac{\tau_c}{u \cdot p}p^{\rho}\nabla_{\rho}\left[\tau_c \dot{f_0}\right],\\
			\nonumber
			\mathcal{C}&=&\Delta^{\mu\nu}_{\alpha \beta}\int_{}^{}dp p^{\alpha}p^{\beta} \frac{\tau_c}{u \cdot p}p^{\rho}\nabla_{\rho} \left[\frac{\tau_c}{u \cdot p}p^{\sigma}\nabla_{\sigma} f_0\right].
		\end{eqnarray}

		\begin{eqnarray}
			\label{eq:a}
			\mathcal{A}&=&-\tau_c \dot{\pi}^{\langle\mu\nu\rangle}-2\tau_c^2 \left(\frac{n_f}{\epsilon+P}
			J_{31}^{(0)-}-J_{31}^{(1)-}\right)\dot{u}^{\langle\mu}\nabla^{\nu\rangle}\alpha
			+\tau_c^2\Delta^{\mu\nu}_{\alpha \beta}J_{31}^{(0)+}\dot{u}^{\beta}
			\left[\frac{\beta qBb^{\alpha \sigma}}{\epsilon+P}V_{\sigma}\right]\nonumber \\ \nonumber
			&&+\tau_c^2\Delta^{\mu\nu}_{\alpha \beta}J_{31}^{(0)+}\dot{u}^{\alpha}\left[\frac{\beta qBb^{\beta \sigma}}
			{\epsilon+P}V_{\sigma}\right]-\tau_c^2 \Delta^{\mu\nu}_{\alpha\beta}J_{31}^{(0)-}\dot{u}^{\beta}\left[\frac{q\beta E^{\alpha}n_f}{\epsilon+P}\right]-\tau_c^2 \Delta^{\mu\nu}_{\alpha\beta}J_{31}^{(0)-}\dot{u}^{\alpha}\left[\frac{q\beta E^{\beta}n_f}{\epsilon+P}\right], \\ \nonumber
			\label{eq:b}
			\nonumber
			\mathcal{B}&=&-2\tau_c^2\left[\left(J_{31}^{(0)+}+J_{42}^{(1)+}\right)\dot{\beta}-\left(J_{31}^{(1)-}+J_{42}^{(2)-}\right)\dot{\alpha}\right]\sigma^{\mu\nu}-2\tau_c^2\nabla^{\langle\mu}\left(\dot{u}^{\nu\rangle}\beta J_{42}^{(1)+}\right), \\ \nonumber
			\label{eq:c}
			\nonumber
			\mathcal{C}&=&2 \nabla^{\langle\mu}\left(\dot{u}^{\nu\rangle}\beta \tau_c^2 J_{42}^{(1)+}\right)
			+2 \nabla^{\langle\mu}\left[\nabla^{\nu\rangle}\alpha \tau_c^2 \left(J_{42}^{(2)-}
			-\frac{1}{h}J_{42}^{(1)-}\right)\right]
			-4\beta \tau_c^2 \left(2J_{63}^{(3)+}+J_{42}^{(1)+}\right)\sigma^{\langle\mu}_{\rho}
			\sigma^{\nu\rangle\rho}\\ \nonumber
			&&-\frac{20}{3} \beta \tau_c^2 J_{42}^{(1)+}\theta \sigma^{\mu\nu} - \frac{28}{3}
			\beta \tau_c^2 J_{63}^{(3)+} \theta \sigma^{\mu\nu}-4 \beta \tau_c^2 \left(J_{42}^{(1)+}+2J_{63}^{(3)+}
			\right) \sigma^{\langle\mu \rho}\omega^{\nu\rangle}_{\rho}\\ 
			&& +2\tau_c^2 \nabla^{\langle\mu}\left[J_{42}^{(1)+}\left(\frac{\beta qBb^{\nu\rangle
					\gamma}V_\gamma}{\epsilon+P}\right)\right]-2\tau_c^2 \nabla^{\langle\mu}\left[J_{42}^{(1)-}\left(\frac{\beta qE^{\nu\rangle}n_f}{\epsilon+P}\right)\right].
		\end{eqnarray}

		The rest of the terms give:
		
		\begin{eqnarray}\nonumber
			\mathcal{I}_{2}&=&q\tau_c^2 \beta J_{31}^{(1)-}\Delta^{\mu\nu}_{\alpha\beta}\left(E^{\alpha}\dot{u}^{\beta}+E^{\beta}\dot{u}^{\alpha}\right)+\nabla^{\langle \mu}\left(q\tau_c^2 \beta E^{\nu \rangle}J_{42}^{(2)-}\right), \\ \nonumber
			\mathcal{I}_{3}&=&\Delta^{\mu\nu}_{\alpha\beta}q\tau_c^2 \left(J_{31}^{(1)-}\left(\dot{\beta}\left(B^{\alpha\beta}+B^{\beta\alpha}\right)+E^{\beta}\nabla^{\alpha}\beta +E^{\alpha}\nabla^{\beta}\beta \right) \right) \nonumber \\
			&&+\Delta^{\mu\nu}_{\alpha\beta}q\tau_c^2 \left( J_{42}^{(2)-}\left(E^{\beta}\nabla^{\alpha}\beta+E^{\alpha}\nabla^{\beta}\beta\right)+\beta J_{41}^{(2)-}E^{\alpha}\dot{u}^{\beta}\right) \nonumber \\
			&&+\Delta^{\mu\nu}_{\alpha\beta}q\tau_c^2 \left(\beta J_{42}^{(2)-}\left(\theta B^{\alpha\beta}+B^{\alpha}_{\nu}\nabla^{\beta}u^{\nu}+B^{\alpha}_{\nu}\nabla^{\nu}u^{\beta} \right) \right) \nonumber\\
			&&+\Delta^{\mu\nu}_{\alpha\beta}q\tau_c^2\left(\beta J_{41}^{(2)-}E^{\beta}\dot{u}^{\alpha}+\beta J_{42}^{(2)-}\left(\theta B^{\beta\alpha}+B^{\beta}_{\nu}\nabla^{\alpha}u^{\nu}+B^{\beta}_{\nu}\nabla^{\nu}u^{\alpha}\right)\right) \nonumber \\
			&&-\Delta^{\mu\nu}_{\alpha\beta}q\tau_c^2\ J_{31}^{(2)-}\left( \dot{\alpha}B^{\alpha\beta}+E^{\alpha}\nabla^{\beta}\alpha +\dot{\alpha}B^{\beta\alpha}+E^{\beta}\nabla^{\alpha}\alpha \right) \nonumber \\ \nonumber
			&&+\Delta^{\mu\nu}_{\alpha\beta}q\tau_c^2 \beta J_{52}^{(3)-}\left( E^{\beta}\dot{u}^{\alpha}+ E^{\alpha}\dot{u}^{\beta}\right)-\Delta^{\mu\nu}_{\alpha\beta}q\tau_c^2  J_{42}^{(3)-}\left(E^{\alpha}\nabla^{\beta}\alpha +E^{\beta}\nabla^{\alpha}\alpha \right), \\ 
			\mathcal{I}_{4}&=&-2\Delta^{\mu\nu}_{\alpha\beta}q^2 \tau_c^2\beta E^{\alpha}E^{\beta}\left(J_{42}^{(3)+}+J_{31}^{(2)+}\right).\nonumber
		\end{eqnarray}

		\section{General expressions of transport coefficients} 
		\renewcommand{\arraystretch}{2}
	\begin{table*}[h!]
\begin{center}
\begin{tabular}{|p{1.5cm}|p{9.0cm}|}
 \hline
 $\tau_{\Pi V}$&$-\beta \frac{\partial}{\partial \beta}\left[\frac{5}{3\beta_V}\left(J_{42}^{(2)-}-\frac{n_f}{\epsilon+P}J_{42}^{(1)-}\right)\right]-\left[\frac{2}{3\beta_{V}} \ \left(\frac{J_{31}^{(0)-}}{h}-J_{31}^{(1)-}\right)-\beta \frac{\partial }{\partial \beta}\left(\frac{5}{3h\beta_{V}}J_{42}^{(1)-}-\frac{5}{3\beta_{V}}   J_{42}^{(2)-}\right)\right]$\\
  \hline
  $\lambda_{\Pi V}$&$\left(\frac{\partial }{\partial \alpha}+h^{-1}\frac{\partial }{\partial \beta}\right)\left[\frac{5}{3\beta_V}\left(J_{42}^{(2)-}-\frac{n_f}{\epsilon+P}J_{42}^{(1)-}\right)\right]-\frac{1}{3\beta \beta_V}\left(5J_{42}^{(3)+} - \frac{5}{h}J_{42}^{(2)-} + 2J_{31}^{(2)+}- \frac{2}{h}J_{31}^{(1)-} \right)-\left(\frac{\partial }{\partial \alpha}+h^{-1}\frac{\partial }{\partial \beta}\right)\left[\frac{5}{3h\beta_{V}}J_{42}^{(1)-}-\frac{5}{3\beta_{V}}   J_{42}^{(2)-}\right]$\\
  \hline
  $l_{\Pi V}$&$\frac{5}{3\beta_V}\left(J_{42}^{(2)-}-\frac{n_f}{\epsilon+P}J_{42}^{(1)-}\right)-\frac{1}{\beta\beta_{V}}\left[\frac{5\beta}{3h}J_{42}^{(1)-}-\frac{5}{3}  \beta J_{42}^{(2)-}\right]$\\
  \hline
 $\lambda_{\Pi V B}$&$\frac{1}{\beta_{V}}\left(\frac{\partial }{\partial \alpha}+h^{-1}\frac{\partial }{\partial \beta}\right)\left[\frac{5J_{42}^{(1)+}\beta}{3\left(\epsilon+P\right)}\right]+\frac{1}{\beta_{V}}\left[\frac{-1 }{3 (\epsilon+p)}\left(5J_{42}^{(2)-}  +2 J_{31}^{(1)-} \right)\right]$\\
  \hline
  $\chi_{\Pi EE}$&$\frac{-\beta}{3}\left(5J_{42}^{(3)+}+2J_{31}^{(2)+}-\frac{5}{h}J_{42}^{(2)+}-\frac{2}{h}J_{31}^{(1)+}\right)$\\
\hline
 \end{tabular}
\\[13pt]
				\caption{ Transport coefficients appearing in bulk-stress equation Eq.~\eqref{eq:2bulkevolutionexp}.}
				\label{table:bulkjmn}
			\end{center}
		\end{table*}

		\renewcommand{\arraystretch}{2}
		\begin{table*}[h!]
			\begin{center}
	\begin{tabular}{|p{2.5cm}|p{12.0cm}|}
\hline
$\lambda_{VV}$&$-\left(1+\frac{2}{\beta_V}\left(\frac{n_f}{\epsilon+P}J_{42}^{(2)+}-J_{42}^{(3)+}\right)-\frac{1}{\beta_{V}}\left\{\ \left(J_{31}^{(2)+} + 4 J_{42}^{(3)+} \right) - \frac{1}{h} \left( 2J_{42}^{(2)+}+J_{31}^{(1)+}  \right) \right\}\right)$ \\
\hline
$\delta_{VV}$&$\frac{4}{3}+\frac{5}{3\beta_V}\left(\frac{n_f J_{42}^{(2)+}}{\epsilon+P}-J_{42}^{(3)+}\right)+\frac{1}{\beta\beta_{V}}\left[\frac{\beta}{h}\left(\frac{4}{3}J_{31}^{(1)+}+\frac{5}{3}J_{42}^{(2)+}\right)\right]	-\frac{1}{\beta\beta_{V}}\left[ \beta \left(\frac{7}{3}J_{31}^{(2)+}+\frac{10}{3}J_{42}^{(3)+}\right)-\left( J_{20}^{(1)+} + J_{21}^{(1)+}\right)\mathcal{X} - \left( J_{20}^{(2)+} + J_{21}^{(2)+}\right)\mathcal{Y}\right]$ \\
\hline
$\delta_{VB}$&$\left(\frac{n_f J_{21}^{(1)-}}{\epsilon+P}-J_{21}^{(2)-}\right)/\beta_{V}+\frac{1}{ \beta_{V}} \left( \frac{J_{21}^{(1)-}}{h} -J_{21}^{(2)-} \right)$\\\hline
$\chi_{VE}$&$\beta \beta_{V}$\\
\hline
$\rho_{VE}$&$-\left(\frac{n_f}{D_{20}}\left[\left(J_{20}^{(0)+}\frac{\partial \chi_{VE}}{\partial \alpha}+J_{10}^{(0)+}\frac{\partial \chi_{VE}}{\partial \beta}\right)h-\left(J_{30}^{(0)+}\frac{\partial \chi_{VE}}{\partial \alpha}+J_{20}^{(0)+}\frac{\partial \chi_{VE}}{\partial \beta}\right)\right]\right)$\\
\hline
 \end{tabular}
 \\[14pt]
				\caption{Transport coefficients appearing in Diffusion evolution equation Eq.~\eqref{eq:fulldiffusion}}
				\label{table:diffusionjmn}
			\end{center}
		\end{table*}

		\renewcommand{\arraystretch}{2}
		\begin{table*}[h!]
			\begin{center}
			\begin{tabular}{|p{1.5cm}|p{13.0cm}|}
 \hline
 $\tau_{\pi V}$&$\beta \frac{\partial}{\partial \beta}\left[\frac{2}{\beta_V }\left(J_{42}^{(2)-}-\frac{n_f}{\epsilon+P}J_{42}^{(1)-}\right)\right]-\frac{2}{\beta_{V}} \left[J_{31}^{(1)-}-\frac{J_{31}^{(0)-}}{h} \right]-\beta \frac{\partial }{\partial \beta}\frac{1}{\chi_{VE}}\left[-\beta J_{42}^{(2)-} + 2 J_{42}^{(1)-}\left(\frac{\beta }{h}\right)\right]$\\
  \hline
  \hline
 $\lambda_{\pi V}$&$\left(\frac{\partial}{\partial \alpha}+h^{-1}\frac{\partial}{\partial \beta}\right)\left[\frac{2}{\beta_V }\left(J_{42}^{(2)-}-\frac{n_f}{\epsilon+P}J_{42}^{(1)-}\right)\right]+\frac{2}{h\beta\beta_{V}}\left(J_{31}^{(1)-}+J_{42}^{(2)-}\right) -\frac{2}{\beta\beta_{V}}\left(J_{31}^{(2)-}+J_{42}^{(3)-}\right)-\left(\frac{\partial }{\partial \alpha}+h^{-1}\frac{\partial }{\partial \beta}\right)\frac{1}{\chi_{VE}}\left[-\beta J_{42}^{(2)-} + 2 J_{42}^{(1)-}\left(\frac{\beta }{h}\right)\right]$\\
  \hline
  \hline
 $l_{\pi V}$&$-\frac{2}{\beta_V }\left(J_{42}^{(2)-}-\frac{n_f}{\epsilon+P}J_{42}^{(1)-}\right)+\frac{1}{\chi_{VE}}\left[-\beta J_{42}^{(2)-} + 2 J_{42}^{(1)-}\left(\frac{\beta }{h}\right)\right]$\\
  \hline
  \hline
 $\lambda_{\pi V B}$&$\frac{1}{\beta_{V}}\left(\frac{\partial }{\partial \alpha}+h^{-1}\frac{\partial }{\partial \beta}\right)\left[{2\beta J_{42}^{(1)+}}/{\left(\epsilon+P\right)}\right]+\frac{1}{\beta \beta_{V}}\left[-\frac{2\beta}{(\epsilon+P)}\left(J_{31}^{(1)-}+J_{42}^{(2)-}\right)\right]$\\
  \hline
  \hline
 $\chi_{\pi EE}$&$2\beta\left(\frac{\left(J_{31}^{(1)-}+J_{42}^{(2)-}\right)}{h} -\left(J_{42}^{(3)+}+J_{31}^{(2)+} \right)\right)$\\
  \hline
 \end{tabular}
\\[14pt]
				\caption{ Transport coefficients appearing in shear-stress evolution equation Eq.~\eqref{eq:2evolutionexp}}
				\label{table:shearjmn}
			\end{center}
		\end{table*}	
	
\newpage
\twocolumngrid
\bibliography{resistive_prd}
\end{document}